\documentclass[
prc,%
10pt,%
final,%
notitlepage,%
oneside,%
twocolumn,%
nobibnotes,%
nofootinbib,
superscriptaddress,%
floatfix,%
floatfix,%
showkeys,%
showpacs]%
{revtex4}
\usepackage{color}
\usepackage{amsfonts}
\usepackage{amsbsy}
\usepackage{mathrsfs}
\usepackage{graphicx}
\def\lsim{\mathrel{\rlap{
\lower4pt\hbox{\hskip-3pt$\sim$}}
    \raise1pt\hbox{$<$}}}     
\def\gsim{\mathrel{\rlap{
\lower4pt\hbox{\hskip-3pt$\sim$}}
    \raise1pt\hbox{$>$}}}     

\begin{document}
\title{
Vortex rings in heavy-ion collisions at energies $\sqrt{s_{NN}}=$ 
3--30 GeV and possibility of their observation} 
\author{Yu. B. Ivanov}\thanks{e-mail: yivanov@theor.jinr.ru}
\affiliation{Bogoliubov Laboratory of Theoretical Physics, JINR, Dubna 141980, Russia}
\affiliation{National Research Nuclear University "MEPhI",  
Moscow 115409, Russia}
\affiliation{National Research Center "Kurchatov Institute",  Moscow 123182, Russia} 
\begin{abstract}
The ring structures that appear in Au+Au collisions at collision energies $\sqrt{s_{NN}}=$ 3 -- 30 GeV are studied.
The calculations are performed within the model of three-fluid dynamics. 
It is demonstrated that a pair of vortex rings are formed, one at forward and another at backward rapidities,
in ultra-central Au+Au collisions at $\sqrt{s_{NN}}>$ 4 GeV. 
The vortex rings carry information about early stage of the collision, 
in particular about the stopping of baryons. It is shown that 
these rings can be detected by measuring the ring observable $R_\Lambda$ even in rapidity range $0<y<0.5$ 
(or $-0.5<y<0$) on the level of 0.5--1.5\% at $\sqrt{s_{NN}}=$ 5 -- 20 GeV.  
At forward/backward rapidities, the $R_\Lambda$ signal is expected to be stronger. 
Possibility of observation of the vortex-ring signal against background of non-collective transverse 
polarization is discussed. 
\end{abstract}
\maketitle

\section{Introduction}

Vortex rings are inherent in fluid dynamics. 
They are developed in a cylindrically symmetric flow of fluid with the
longitudinal velocity depending on the radius. 
Such flow results in formation of toroidal vorticity structures, i.e. the vortex rings. 
An example of such vortex rings is the smoke rings.

Formation of vortex rings in heavy ion collisions at high collision energies, $\sqrt{s_{NN}}=$ 40--200 GeV, 
was predicted in hydrodynamic \cite{Ivanov:2018eej} and transport \cite{Xia:2018tes,Wei:2018zfb} simulations.
Later, the vortex rings were reported at lower energy of 
7.7 GeV in simulations in Refs. \cite{Ivanov:2019wzg,Zinchenko:2020bbc}. 
Earlier, ring-like structures, i.e. half rings, were noticed in semi-central Au+Au collisions 
at even lower energy of 5 GeV \cite{Baznat:2013zx,Baznat:2015eca}. 
The authors of  Refs. \cite{Baznat:2013zx,Baznat:2015eca} called this 
specific toroidal structure as a femto-vortex sheet. 
In recent paper \cite{Tsegelnik:2022eoz}, formation of the vortex rings was predicted in 
Au+Au collisions at $\sqrt{s_{NN}}=$ 7.7--11.5 GeV and even at 4.5 GeV, 
where the ring structure turns out to be more diffuse.

\begin{figure}[!htb]
\includegraphics[width=5.cm]{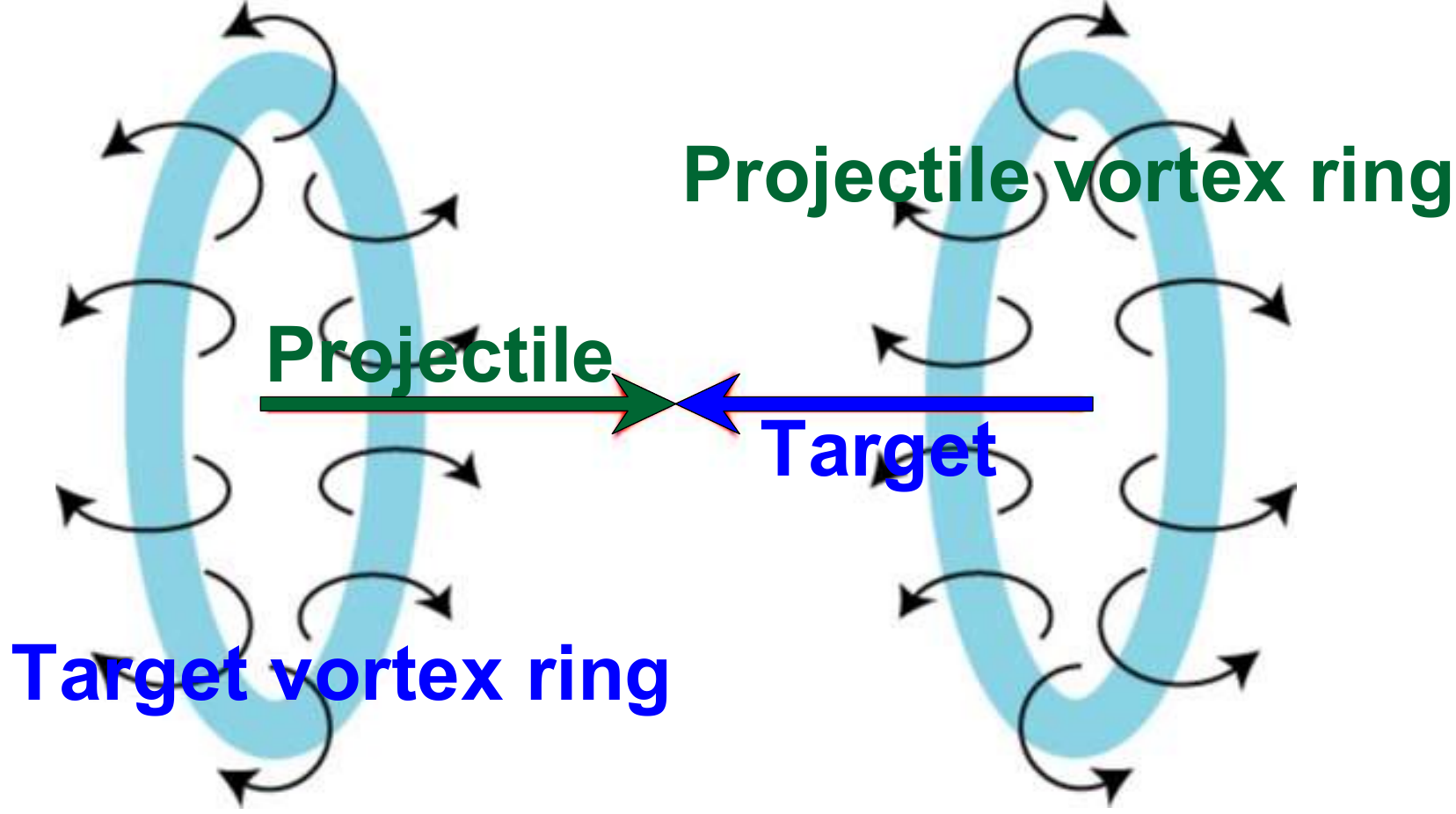}
 \caption{(Color online)
Schematic picture of the vortex rings at forward/backward rapidities. 
Curled arrows indicate direction of circulation of the matter. 
}
\label{fig0}
\end{figure}

Formation of vortex rings is a consequence of partial transparency of colliding nuclei at high energies. 
The matter in the central region is more strongly decelerated
because of thicker matter in the center than that at the periphery.
Therefore, two vortex rings are formed at the periphery of 
the stronger stopped matter in the central region,  
one at forward rapidities and another at backward rapidities.
The peripheral matter acquires a rotational motion. 
Matter rotation is opposite in these two rings. 
A schematic picture of the vortex rings is presented in Fig. \ref{fig0}.

The partial transparency takes place at the early stage of the collision, even before equilibration 
of the produced matter. It determines the strength of vorticity in the vortex rings. 
Therefore, the vortex rings carry information about this early stage of the collision.

In Ref. \cite{Lisa:2021zkj},  the analysis of the toroidal vortex structures was extended to  
proton-nucleus collisions. 
It was predicted that vortex rings are created 
in such collisions at the energy of $\sqrt{s_{NN}}=$ 200 GeV. 
Vortex rings produced by jets propagating through the quark-gluon matter were considered in Ref. 
\cite{Serenone:2021zef}. 
The above predictions 
\cite{Ivanov:2018eej,Xia:2018tes,Wei:2018zfb,Ivanov:2019wzg,Zinchenko:2020bbc,Baznat:2013zx,Baznat:2015eca,Tsegelnik:2022eoz,Lisa:2021zkj,Serenone:2021zef} were obtained within different models. 
Thus, it looks like the vortex rings 
are quite common for the high-energy nucleus(proton)-nucleus collisions. The question is how to observe them.

Authors of Refs. \cite{Lisa:2021zkj,Serenone:2021zef} suggested a ring observable 
\begin{equation}
\label{R_Lambda}
    R_\Lambda (y) =
    \left\langle\frac{{\bf P}_{\Lambda}\cdot\left({\bf e}_z \times {\bf p}\right)}%
{|{\bf e}_z \times {\bf p}|}\right\rangle_y ,
\end{equation}
where ${\bf P}_{\Lambda}({\bf p})$ is the polarization of the $\Lambda$ hyperon, 
${\bf p}$ is its spacial momentum, and ${\bf e}_z$ is the unit vector along the beam, i.e. 
along $z$ axis. Averaging $\left\langle ...\right\rangle_y$ runs over all momenta with fixed rapidity $y$. 
As argued in Ref. \cite{Lisa:2021zkj,Serenone:2021zef}, 
the ring structure may be quantified by means of $R_\Lambda$.

However, the same ring observable turns out to be nonzero in proton-proton and proton-nucleus collisions,
see, e.g.,  Refs. \cite{COSY-TOF:2016vhv,HADES:2014ttv} where a brief survey of earlier experiments is 
also presented. It is referred as a transverse polarization in those experiments. 
At least in proton-proton reactions, the nonzero $R_\Lambda$ is related to the 
correlation of the produced $\Lambda$ with beam direction rather than to the collective ring structure. 
To be precise, only the collective contribution to $R_\Lambda$ due to the  vortex rings
are estimated below 
while the discussion of the background of direct  $\Lambda$ production is postponed to the end of the paper.

This ring observable was applied to analysis of ultra-central Au+Au collisions at
$\sqrt{s_{NN}}=$ 200 GeV  \cite{Lisa:2021zkj}. It was found that nonzero values of $R_\Lambda$
appear only at rapidities $|y|>$ 4, i.e. far beyond the midrapidity window accessible in collider 
experiments. Therefore, questions arise:  
\\
(i) whether the vortex rings can be observed at lower-energy 
collider experiments at BES RHIC (Beam Energy Scan program at the Relativistic
Heavy Ion Collider) and NICA (Nuclotron-based Ion Collider fAcility) 
within the experimental midrapidity window and
\\
(ii) whether the vortex rings are formed in lower-energy collisions 
of the STAR fixed-target program (FXT-STAR) at RHIC 
the forthcoming experiments at the Facility for Antiproton and Ion Research (FAIR), 
where measurements at backward rapidities are possible?

In the present paper, the ring structures in Au+Au collisions at collision energies  
$\sqrt{s_{NN}}=$ 3--30 GeV are studied and  the resulting ring observable is estimated. 
The calculations are performed within the model of the
three-fluid dynamics (3FD) \cite{3FD}.
The 3FD approximation is a minimal way to simulate
the early, nonequilibrium stage of the produced strongly interacting matter. 
It takes into account counterstreaming of the leading baryon-rich matter at the early stage of
nuclear collisions. This counterstreaming results in formation of the vortex rings.

The simulations are done with two different equations
of state (EoS’s): two versions of the EoS with the deconfinement transition \cite{Toneev06},
i.e. a first-order phase transition (1PT) and a crossover
one. The physical input of the present 3FD calculations
is described in Ref.~\cite{Ivanov:2013wha}.

\section{Vortical Rings in Central Collisions}
\label{Vortical Rings}

\begin{figure*}[!htb]
\includegraphics[width=\textwidth]{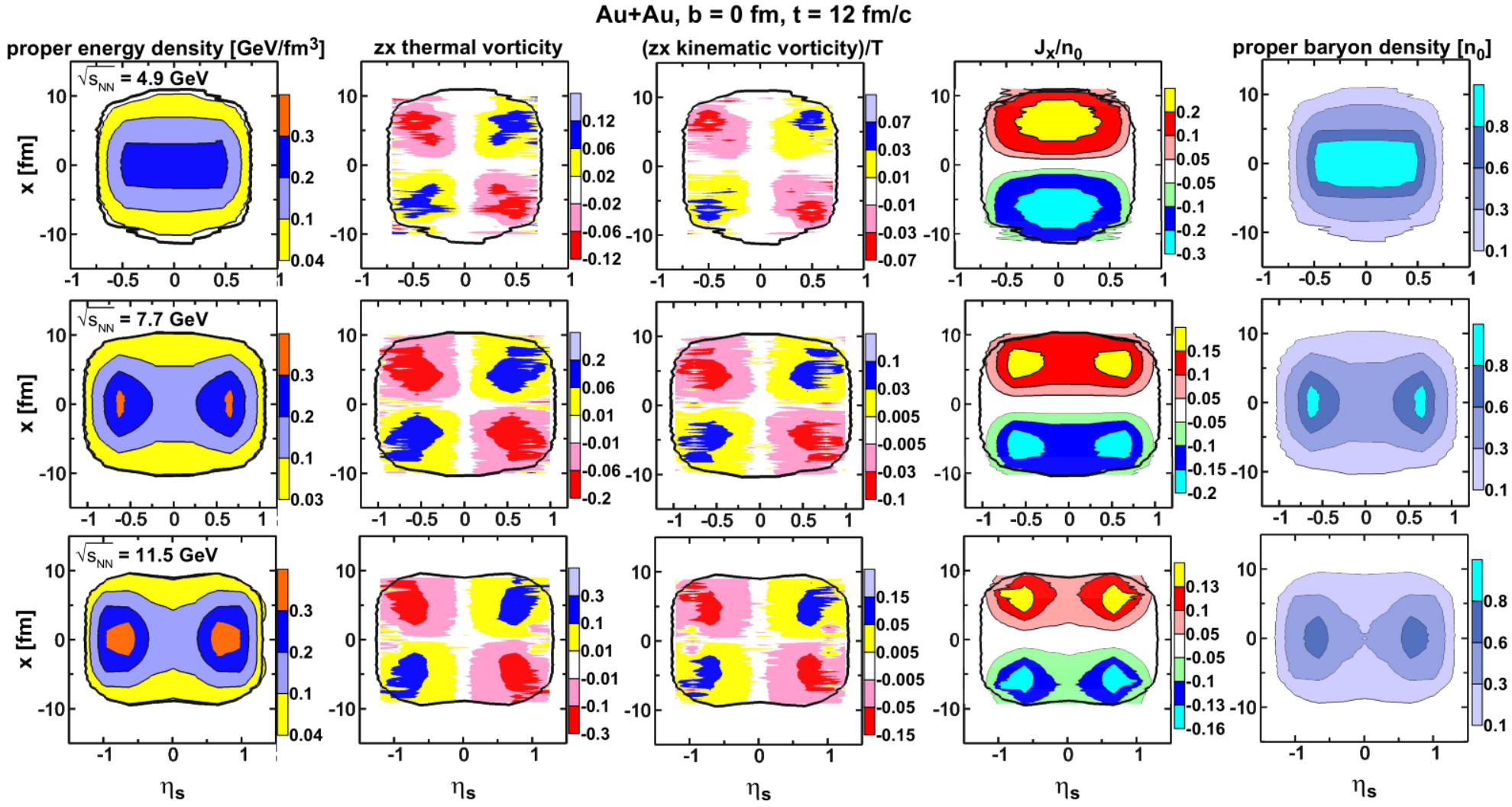}
 \caption{(Color online)
Columns from left to right: the proper energy density (GeV/fm$^3$), 
the proper-energy-density weighted 
thermal $zx$ vorticity, similarly weighted kinematic $zx$ vorticity divided by temperature ($T$), 
$x$-component of the baryon current ($J_x$) in units of normal nuclear density ($n_0=0.15$ 1/fm$^3$), 
and the proper baryon density ($n_B$)  in units of $n_0$
in the $x\eta_s$  plane at time instant $t=$ 12 fm/c 
in the ultra-central ($b=$ 0 fm) Au+Au collisions at $\sqrt{s_{NN}}=$ 4.9--11.5 GeV. 
$\eta_s$  is the space-time rapidity along the beam ($z$ axis) direction.
Calculations are done with the crossover EoS. 
The bold solid contour displays the border of nuclear matter with $n_B> 0.1 n_0$. 
}
\label{fig1}
\end{figure*}

The particle polarization is treated within the  thermodynamic approach \cite{Becattini:2013fla}, in which
it is related to the thermal vorticity
   \begin{eqnarray}
   \label{therm.vort.}
   \varpi_{\mu\nu} = \frac{1}{2}
   (\partial_{\nu} \beta_{\mu} - \partial_{\mu} \beta_{\nu}), 
   \end{eqnarray}
where  $\beta_{\mu}=u_{\mu}/T$ 
with $u_{\mu}$ and $T$ being the local collective velocity of the matter and its 
 temperature, respectively.  
The thermal vorticity is directly related to the mean spin vector of spin
1/2 particles with four-momentum $p$, produced around
point $x$ on freeze-out hypersurface
   \begin{eqnarray}
\label{xp-pol}
 S^\mu(x,p)
 =\frac{1}{8m}     [1-n_F(x,p)] \: p_\sigma \epsilon^{\mu\nu\rho\sigma} 
  \varpi_{\rho\nu}(x), 
   \end{eqnarray}
where $n_F(x,p)$ is the Fermi-Dirac distribution function and $m$ is mass of the 
considered particle. 
The polarization vector of $S$-spin particle is defined as $P^\mu_{S} = S^\mu /S$. 
The polarization of the $\Lambda$ hyperon is measured in its rest frame, therefore the 
$\Lambda$ polarization should be additionally boosted to the $\Lambda$ rest frame.

Let us first consider the structure of the thermal-vorticity field in 
heavy-ion collisions in the $x\eta_s$ plane,  
where $\eta_s = (1/2) \ln\left[(t+z)/(t-z)\right]$
is the longitudinal space-time rapidity and $z$ is the coordinate along the beam direction. 
The advantage of this $\eta_s$ is that it is 
equal to the kinematic longitudinal rapidity 
defined in terms of the particle momenta in the self-similar one-dimensional expansion 
of the system.

The plot of $\varpi_{zx}$ in ultra-central ($b=0$ fm) Au+Au collisions at $\sqrt{s_{NN}}=$ 4.9--11.5 GeV 
in the $x\eta_s$ plane is presented in Fig. \ref{fig1}.  
In order to suppress contributions of almost empty regions, 
the displayed thermal-vorticity $\varpi_{zx}$ is averaged with the weight of 
proper energy density (also presented in Fig. \ref{fig1}) 
similarly to that in Refs. \cite{Ivanov:2018eej,Ivanov:2019wzg}. 
The $x\eta_s$ plane would be a reaction plane in case of non-central collisions. 
In our case ($b=0$ fm), all the planes passing through the $z$ axis are equivalent because of the axial symmetry. 
These are the plots at time instants close to the freeze-out. 
In order to see correlations of the thermal-vorticity with other quantities, 
plots of the proper energy density, 
kinematic $zx$ vorticity     
$$\omega_{\mu\nu} = (1/2) (\partial_{\nu} u_{\mu} - \partial_{\mu} u_{\nu}),$$ 
divided by temperature,
$x$-component of the baryon current ($J_x$) and 
the proper baryon density are also displayed.

As seen, the thermal-vorticity reveals a ring structure similar to that 
schematically displayed in Fig. \ref{fig0}.
The $x\eta_s$ plane is a cut 
of these rings by the plane passing through the axis of these rings. 
This ring structure is seen even at $\sqrt{s_{NN}}=$ 4.9 GeV. 
The kinematic $zx$ vorticity (the third column of panels in Fig. \ref{fig1}) 
reveals the same ring structure. This indicates that these rings are due to the incomplete  
stopping of the peripheral parts of the colliding nuclei. If the rings were formed as a result of 
the hydrodynamic quasi-one-dimensional expansion of initially stopped matter, 
then the sign of $\omega_{zx}/T$ would be opposite because the longitudinal flow velocity 
would decrease from the center of the flow to its periphery, as in the case of
high-energy proton-nucleus collisions \cite{Lisa:2021zkj}. 
Comparing the scales of $\varpi_{zx}$ and $\omega_{zx}/T$ in Fig. \ref{fig1}, 
we see that approximately half of the magnitude of the thermal vorticity results from 
derivatives of the inverse temperature.

As seen from Fig. \ref{fig1}, these thermal-vorticity rings correlate with transverse 
component of the baryon current ($J_x$). It means that the vortical rings expand, 
which is important for their observation. At the same time, the proper energy and density 
distributions reveal a disk rather than ring structure although at the same  
space-time rapidities. At 4.9 GeV, this is already a central fireball rather than two disks.

At considered moderately relativistic energies, the expansion dynamics of the system  
is not of the self-similar one-dimensional character. Therefore, the space-time rapidity $\eta_s$ 
is not equal to the kinematic longitudinal rapidity in terms of the particle momenta. 
Nevertheless, it can be used to approximately estimate the rapidity location of these vortex rings. 
As seen from Fig. \ref{fig1}, at 4.9 GeV these rings are located slightly below $|\eta_s|=$ 0.5 and 
at $|\eta_s|\approx$ 0.5, if $\sqrt{s_{NN}}=$ 7.7 GeV. This location does not restrict their observation
because the FXT-STAR experiments allow measurements at very backward
rapidities. At $\sqrt{s_{NN}}=$ 11.5 GeV, where the estimation in terms of $\eta_s$ is more reliable, 
these vortex rings are already located at $|\eta_s|\approx$ 0.5--1.0, which is slightly beyond the 
rapidity window of the collider experiment. Nevertheless, the inner parts of these rings are still 
at $|\eta_s|<$ 0.5. Therefore, their effect can be observed.

\section{Ring Observable}
\label{ring observable}

To quantify the above qualitative considerations, let us turn to the ring observable
of Eq. (\ref{R_Lambda}). Our goal is to estimate the expected ring observable
in the ultra-central ($b =$ 0 fm) Au+Au collisions rather than to calculate it. 

In terms of hydrodynamic quantities, the contribution of an element of the freeze-out surface 
$d \Sigma_{\nu}$ to the ring observable reads
\begin{eqnarray}
\label{eq:Rlambda}
    R_\Lambda (x,p) =
    \frac{\varepsilon^{\mu\nu\rho\sigma}P_{\mu}^\Lambda n_{\nu}  e_{\rho} p_{\sigma}}
    {|\varepsilon^{\mu\nu\rho\sigma}n_{\nu} e_{\rho}p_{\sigma}|} \quad . 
\end{eqnarray} 
where $e^{\sigma}=(0,0,0,1)$ is the unit vector along the beam ($z$) axis, $n_{\nu}$
is the normal vector to the element of the freeze-out surface.  
This expression coincides with Eq. (10) in Ref. \cite{Lisa:2021zkj}.  

To calculate the ring observable $R_\Lambda (y)$, $R_\Lambda (x,p)$  should be averaged
over the whole freeze-out surface $\Sigma$ and particle momenta
   \begin{eqnarray}
\label{polint}
 R_{\Lambda} (y_h)
 = \frac{\int (d^3 p/p^0) \int_{\Sigma (y_h)} d \Sigma_\lambda p^\lambda
n_{\Lambda} (x,p)   R_{\Lambda} (x,p)}%
 {\int (d^3 p/p^0) \int_{\Sigma (y_h)} d\Sigma_\lambda p^\lambda \, n_{\Lambda} (x,p)}, 
   \end{eqnarray}
where $n_{\Lambda}$ is the distribution function of $\Lambda$'s. 
Here the approximation has already been made: the constraint of 
fixed rapidity ($y$) imposed on the momentum integration is replaced by that of 
fixed hydrodynamical rapidity 
   \begin{eqnarray}
   \label{yh}
y_h = \frac{1}{2} \ln \frac{u^0+u^3}{u^0-u^3} , 
   \end{eqnarray}
based on hydrodynamical 4-velocity $u^\mu$, similarly to that in Ref.  \cite{Ivanov:2022ble}. 
The $y_h$ constraint is imposed on the 
freeze-out surface integration and is denoted as $\Sigma (y_h)$ in Eq. 
(\ref{polint}). 
At moderately relativistic energies,
the hydrodynamical rapidity is a more reliable approximation to the true rapidity than 
the space-time rapidity 
$\eta_s$. 

Let us further proceed with approximations. Similarly to that in Ref.  \cite{Ivanov:2022ble}, let us use a 
simplified version of the freeze-out, i.e. an isochronous one that implies 
$n_{\nu}=(1,0,0,0)$ and $(d^3 p/p^0) d \Sigma_\lambda p^\lambda = d^3 p \: d^3 x$. 
The freeze-out instant is associated with time,
when the energy density averaged over the central
region reaches the value deduced from the conventional 3FD freeze-out. 
In conventional 3FD simulations, a differential,
i.e. cell-by-cell, freeze-out is implemented \cite{Russkikh:2006aa}.

Expression (\ref{xp-pol}) for $S^\mu_{\Lambda}(x,p)$ is also simplified. 
The factor $(1-n_{\Lambda})\approx 1$ is taken because the $\Lambda$ production takes
place only in high-temperature regions, where Boltzmann
statistics dominates. As spatial components of $\varpi_{\rho\nu}(x)$ are of the prime interest, 
the approximation  $p_\sigma \epsilon^{\mu\nu\rho\sigma} \varpi_{\rho\nu} \approx
p_0 \epsilon^{\mu\nu\rho 0} \varpi_{\rho\nu} \approx
m_{\Lambda} \epsilon^{\mu\nu\rho 0} \varpi_{\rho\nu}$ is made. 
The latter approximation, $p_0\approx m_{\Lambda}$,    
reduces $S^\mu_{\Lambda}(x,p)$, which is quite suitable for the purpose of  
upper estimate of the ring observable. 
The boost of $S^\mu_{\Lambda}(x,p)$ is neglected, which also reduces $S^\mu_{\Lambda}(x,p)$. 
After application of all these approximations, $S^\mu_{\Lambda}$, and hence  
$P^\mu_{\Lambda}$, becomes momentum independent. Therefore,  
the momentum averaging in Eq. (\ref{polint})
can be performed first, leaving $P^\mu_{\Lambda}$
beyond the scope of this averaging:
   \begin{eqnarray}
\label{polint-dx}
 R_{\Lambda} (y_h)
&\approx &
 \int_{\Sigma (y_h)} d^3 x \, \rho_{\Lambda} (x) 
    \frac{{\bf P}_{\Lambda}\cdot\left({\bf u} \times {\bf e}_z\right)}%
{({\bf u}_T^2 + 2  T/m_{\Lambda})^{1/2}}
\cr
&/&
{\int_{\Sigma (y_h)} d^3 x \rho_{\Lambda} (x)}, 
   \end{eqnarray}
where $\rho_{\Lambda} (x)$ is the density of $\Lambda$'s, ${\bf u}_T$ is transverse component of the 
fluid velocity, and $T$ is the temperature. This expression is obtained in the non-relativistic 
approximation for the transverse collective motion. Indeed, at the freeze-out 
$T\approx$ 100 MeV $\ll m_{\Lambda}$ and $v_T \lsim$ 0.4 at midrapidity \cite{STAR:2017sal}.
At the forward/backward rapidities considered here, these values are even smaller. 
Here $({\bf u}_T^2 + 2  T/m_{\Lambda})^{1/2}$ stands for $\left\langle |{\bf p}_T|\right\rangle/m_{\Lambda}$.

It is important that the system together with the vortex rings radially expands at the freeze-out stage. 
This expansion even determines the sign of $R_{\Lambda}$, as seen from Eq. (\ref{polint-dx}). 
The contributions to $R_{\Lambda}$ from particles emitted along the ${\bf u}$ 
(more precisely, with momenta ${\bf p}_T\cdot{\bf u}_T>0$) and 
in opposite to ${\bf u}$ directions (i.e. ${\bf p}_T\cdot{\bf u}_T<0$)
partially cancel each other. 
The effect of this partial cancellation is described by the 
$({\bf u}_T^2 + 2  T/m_{\Lambda})^{1/2}$ denominator in Eq. (\ref{polint-dx}). 
This cancellation is negligible if the speed of the radial expansion is much larger than the 
thermal velocity of the fluid constituents.

Based on the axial symmetry of the ultra-central ($b=$ 0) collisions, 
averaging in Eq. (\ref{polint-dx})  can be restricted  
by the quadrant ($x>0,z>0$) of the ``reaction'' plane $xz$, 
[the ($x>0,\eta_s>0$) quadrant in Fig. \ref{fig1}], if 
$y_h>0$. For negative $y_h$, it is  ($x>0,z<0$) quadrant. 
Thus, this averaging becomes identical to that done in Ref. \cite{Ivanov:2022ble}
for calculation of the global polarization but in a restricted region, i.e. 
the ($x>0,z>0$) quadrant. The only difference is that  the 
averaging over cells runs with additional weight 
$x \; u_x/(u_x^2 + 2  T/m_{\Lambda})^{1/2}$, 
where $x$ takes into account that the integration runs over $rdr\: dz$ in cylindrical 
coordinates at fixed azimuthal angle.

\begin{figure}[!htb]
\includegraphics[width=6.5cm]{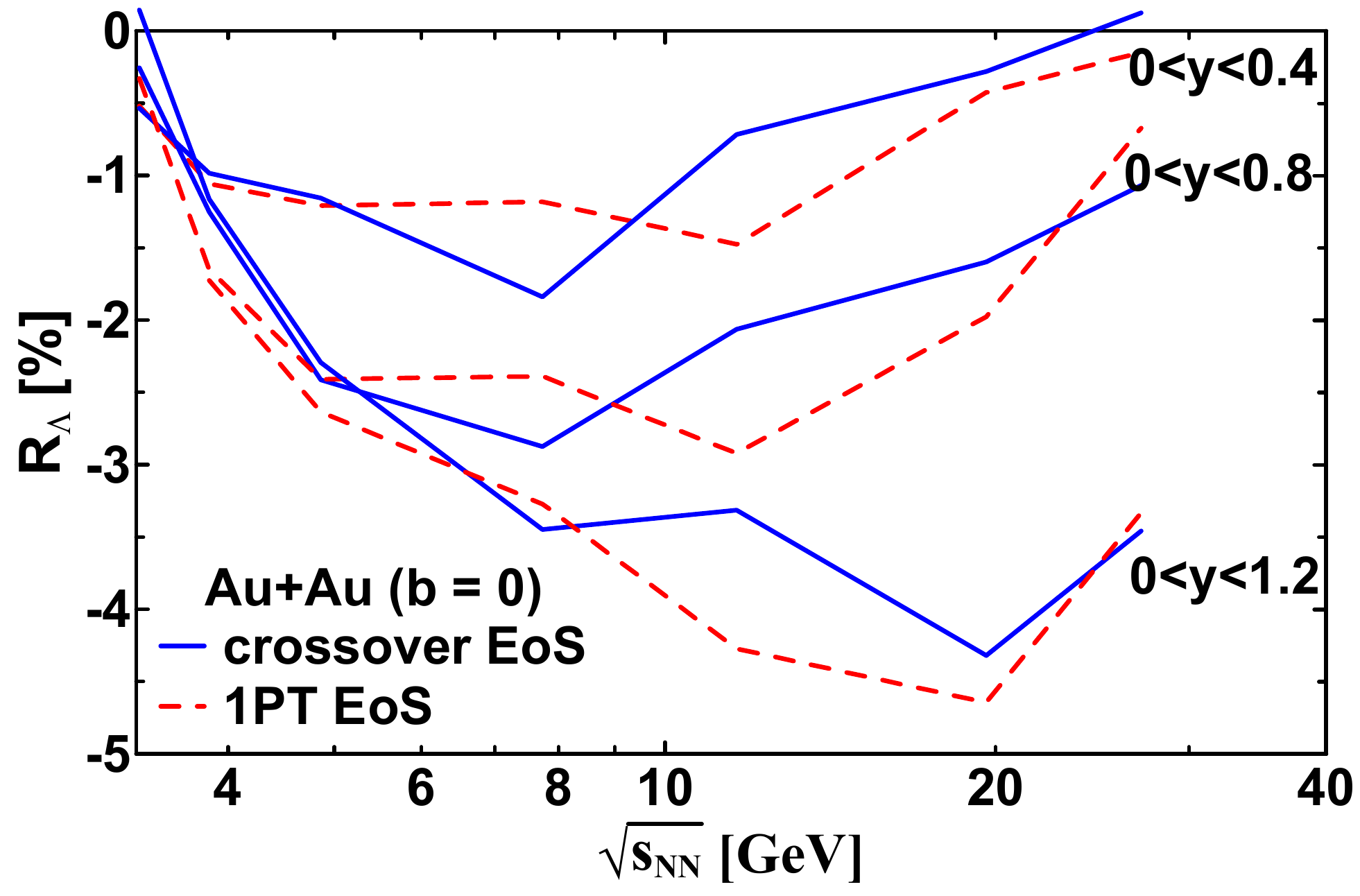}
 \caption{(Color online)
The $R_\Lambda$ quantity,  
averaged over different rapidity ranges,
in the ultra-central Au+Au collisions as function of $\sqrt{s_{NN}}$. 
Calculations are done with the crossover and 1PT EoS's. 
}
\label{fig2}
\end{figure}

\begin{figure}[!htb]
\includegraphics[width=6.5cm]{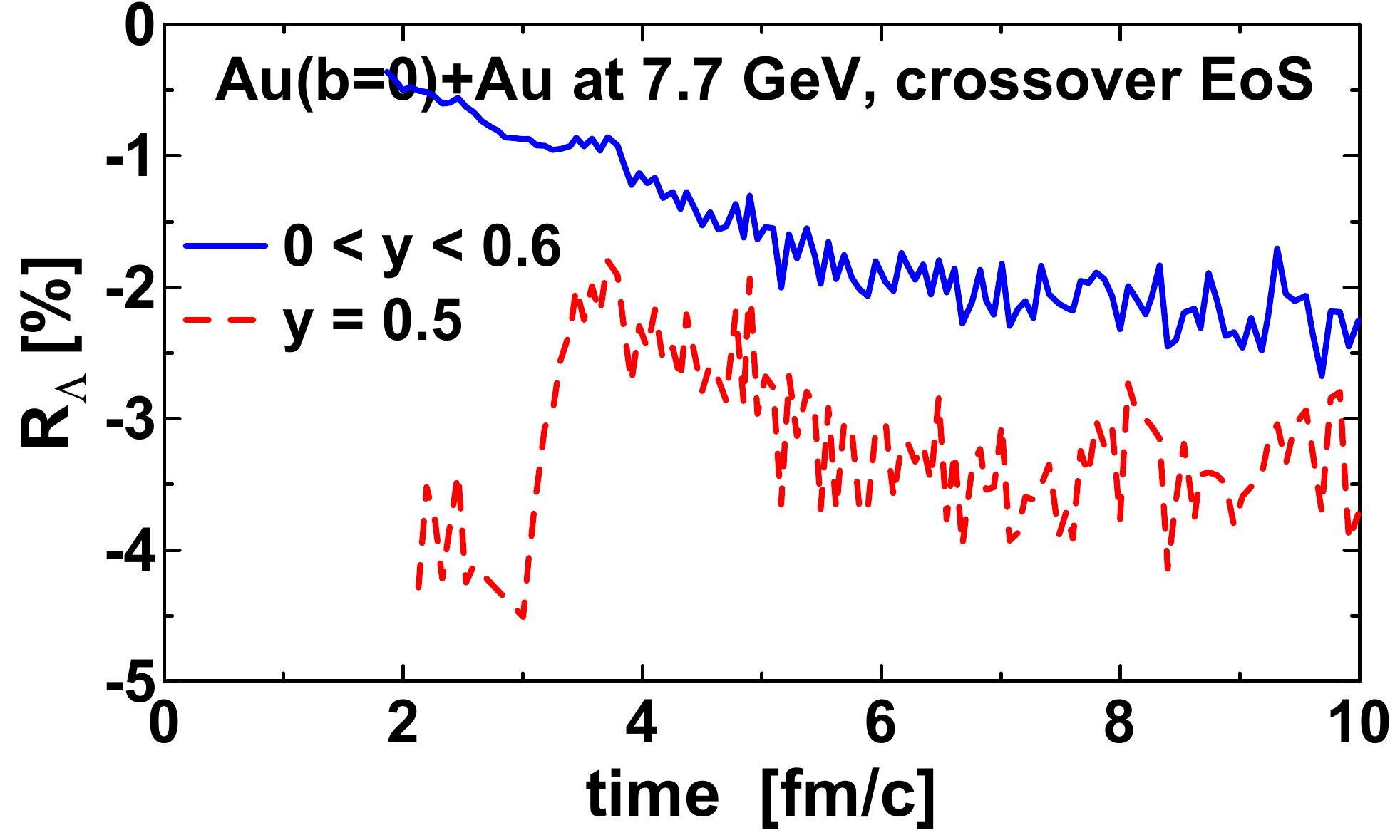}
 \caption{(Color online)
Time dependence of the $R_\Lambda$ quantity,  
averaged over rapidity range of $0<y<0.6$ (solid line) and at $y=$ 0.5 (dashed line),
in the ultra-central ($b=$ 0 fm) Au+Au collisions at $\sqrt{s_{NN}}=$ 7.7 GeV.  
Calculations are done with the crossover EoS. 
}
\label{fig3}
\end{figure}

The estimation of the $R_\Lambda$ quantity,  
averaged over different rapidity ranges,
in the ultra-central ($b=$ 0) Au+Au collisions as function of $\sqrt{s_{NN}}$ is displayed 
in Fig.  \ref{fig2}. For definiteness, all these rapidity ranges are located at positive rapidities because 
$R_{\Lambda}(y)$ is odd function of $y$. 
In fixed rapidity region, the magnitude of $R_{\Lambda}$ first increases with the collision energy rise, reaches a maximum, 
the position and height of which depend on the rapidity window, and then decreases.  
The height of the maximum is larger in more wide rapidity windows. All these features are expected from 
the above qualitative consideration.  
The presented calculations have numerical uncertainty of 10--15\%, 
as seen from Fig.  \ref{fig3} where the time dependence of the $R_\Lambda$ quantity 
is displayed.
These numerical fluctuations result from averaging over very restricted spacial region 
rather than the whole volume of the system, as 
it is done for the global polarization \cite{Ivanov:2022ble}. These fluctuations are larger for 
$R_{\Lambda}$ at fixed $y$, also shown in Fig.  \ref{fig3}, because the 
averaging runs over even more restricted spacial region in this case. 

Calculations are done with the crossover and 1PT EoS's. 
Predictions of the different EoS's differ, but this difference is comparable with numerical 
uncertainty at energies below 7 GeV and above 12 GeV.
However, at 7 $\lsim\sqrt{s_{NN}}\lsim$ 12 GeV this difference exceeds the numerical uncertainty. 
This energy range correlates with that of the earlier predicted irregularity 
in the excitation function of the baryon stopping \cite{Ivanov:2013wha,Ivanov:2012bh},
where results of different EoS's also differ. 
It is not surprising because the incomplete baryon stopping (or partial transparency) is 
the driving forth the vortex-ring formation. 
In particular, 
this is the reason of correlation between the thermal vorticity and baryon current, see Fig. \ref{fig1}.

\begin{figure}[!htb]
\includegraphics[width=8.6cm]{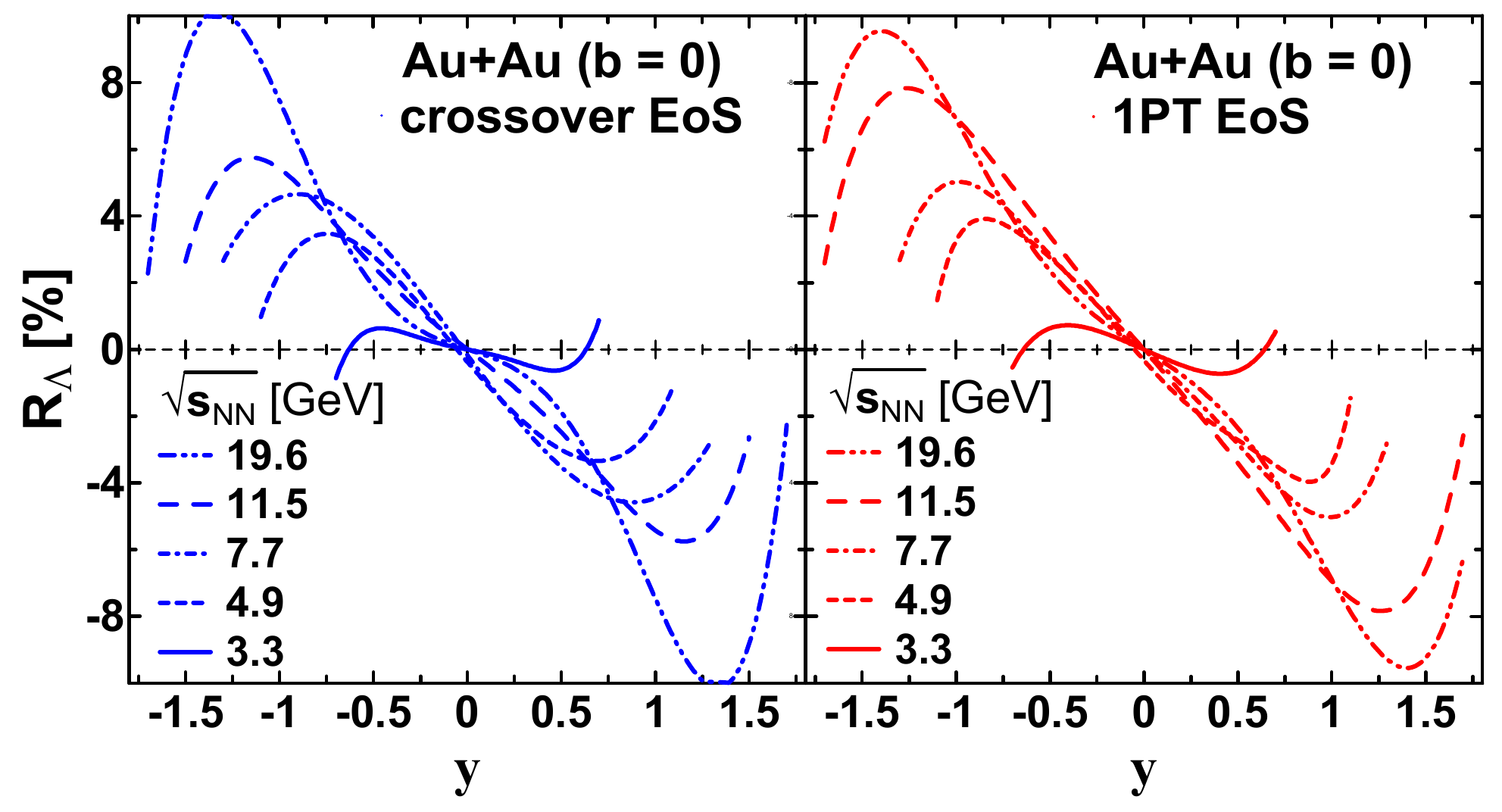}
 \caption{(Color online)
Rapidity dependence of the  $R_\Lambda$ quantity
in the ultra-central ($b=$ 0) Au+Au collisions at $\sqrt{s_{NN}}=$ 3.3--19.6 GeV. 
Calculations are done with the crossover (left panel) and 1PT (right panel) EoS's. 
}
\label{fig4}
\end{figure}

Rapidity dependence of the $R_\Lambda$ quantity 
in the ultra-central Au+Au collisions at different $\sqrt{s_{NN}}$ 
is shown in Fig.  \ref{fig4}. 
In view of the large numerical fluctuations, cf. Fig.  \ref{fig3}, 
the lines in Fig.  \ref{fig4} are smoothed by means of 
orthogonal-polynomial fit. 
Again, as expected, extreme values of $R_{\Lambda}$ are reached at 
forward/backward rapidities, where the vortex rings are located. 
Positions of these extreme values move to larger $|y|$ with the collision energy rise. 
However, even at 7.7 $<\sqrt{s_{NN}}<$ 19.6 GeV, 
sizable values of $R_{\Lambda}$ can be expected at $|y|=$ 0.5 attainable in the collider mode.    
At lower energies, $\sqrt{s_{NN}}\leq$ 7.7 GeV, measurements at backward rapidities are possible 
in the FXT-STAR and future FAIR experiments. At $\sqrt{s_{NN}}=$ 7.7 GeV, $R_{\Lambda}$ can reach 
values of 5\% at backward rapidities. 
Feed-down from higher-lying resonances ($\Sigma^*$ and $\Sigma^0$) in $R_{\Lambda}$ has been taken into account 
similarly to that done for the global polarization in Ref. \cite{Ivanov:2022ble}.
This feed-down reduces $R_\Lambda$ by 
10--15\%.

\section{Summary and Discussion}
\label{Summary}

The ring structures that appear in Au+Au collisions at energies $\sqrt{s_{NN}}=$ 3 -- 30 GeV were studied, 
and the resulting ring observable was estimated.
The calculations were performed within the 3FD model \cite{3FD}. 
It was demonstrated that a pair of vortex rings are formed, one at forward and another at backward rapidities,
in ultra-central Au+Au collisions at $\sqrt{s_{NN}}\gsim$ 4 GeV. 
The matter rotation is opposite in these two rings. They are 
formed because at the early stage of the collision the matter in the vicinity of the beam axis is stronger decelerated 
than that at the periphery. 
Thus, the vortex rings carry information about this early stage, in particular about the stopping of baryons.

Such 3FD dynamics with the incomplete stopping turned out to be successful in describing the global $\Lambda$ polarization 
\cite{Ivanov:2022ble,Ivanov:2020udj}, bulk \cite{Ivanov:2013wha,Ivanov:2013yqa} 
and flow \cite{Ivanov:2014ioa,Ivanov:2014zqa}   observables
at moderately relativistic energies. Therefore, the present predictions of the vortex rings have a solid background.

These rings can be detected by measuring the ring observable $R_\Lambda$ even in rapidity range $0<y<0.5$ 
(or $-0.5<y<0$).  The $R_\Lambda$ signal is stronger in wider rapidity ranges. 
For instance, magnitude of the ring observable may reach values of 2--3\% at $\sqrt{s_{NN}}=$ 5--20 GeV, 
if rapidity window is extended to $0<y<0.8$. Measurements in fixed-target experiments, such as 
FXT-STAR and the forthcoming experiments at FAIR, give additional advantage. They allow measurements  
at backward rapidities, where the $R_\Lambda$ signal is expected to be more pronounced.

Only ultra-central Au+Au collisions were considered in this paper because the axial symmetry of the system
makes the  $R_\Lambda$ estimation easier in this case.
Asymmetric vortex rings are also formed 
in semi-central collisions at $\sqrt{s_{NN}}>$ 5 GeV. 
The corresponding $R_\Lambda$ polarization should be asymmetric in the reaction plain.

The $R_\Lambda$ quantity also contains contribution 
of direct $\Lambda$ production with the $\Lambda$ polarization  
correlated with beam direction. 
In heavy-ion collisions this type of polarization
is expected to be diluted due to rescatterings in the medium  \cite{Panagiotou:1986zq,Ayala:2001jp}.
A vanishing $\Lambda$
polarization has been proposed as a possible signature for the formation
of a Quark GIuon Plasma (QGP) in relativistic heavy-ion collisions \cite{Panagiotou:1986zq}.
This prediction was one of motivations of the 
experimental study of the transverse polarization of $\Lambda$ hyperons 
produced in Au+Au collisions (10.7$A$ GeV) 
\cite{Bellwied:2002cz}.
The result revealed no significant differences of the polarization from those observed
in $pp$ and $pA$ collisions. 
However, the question of the nature of the obtained polarization remains open, 
i.e., which part of it results from the direct $\Lambda$ production  
and which part, from to vortex rings.

Different transverse-momentum dependence may be used to distinguish these two mechanisms of the 
transverse polarization. While the magnitude of the transverse polarization due to the direct $\Lambda$ production
linearly rises with $\Lambda$ transverse momentum, low-$p_T$ $\Lambda$'s should dominate in 
the $R_\Lambda$ due to vortex rings because these rings are collective phenomena.  
The $p_T$ dependence of the vortex-ring $R_\Lambda$ is expectred to be decreasing, similarly to that for   
the global polarization \cite{Okubo:2021dbt,STAR:2021beb,Adams:2021idn}. 
Therefore, limiting the transverse momentum from above would enhance the contribution of the vortex rings.

Any case, the calculations of the $R_\Lambda$ due to vortex rings should be complemented by 
simulations of the $\Lambda$ transverse polarization due to the direct $\Lambda$ production, 
similarly to that done for the ultra-central Au+Au collisions at $\sqrt{s_{NN}}=$ 9 GeV
in Ref. \cite{Nazarova:2021lry}.  
That simulation disregarded the polarization dilution due to rescatterings in the medium.
Therefore it gave the upper limit ($\approx -5\%$ ) on  magnitude of the
mean transverse $\Lambda$ polarization. For the practical use, such simulations should  
take into account this dilution \cite{Ayala:2001jp}.

The transverse $\bar{\Lambda}$ polarization measured in $pp$ collisions so far 
is consistent with zero \cite{ATLAS:2014ona}. 
On the other hand, the collective fluid vorticity in $AA$ collisions 
polarizes all emitted particles and hence  
$R_\Lambda\approx R_{\bar{\Lambda}}$ should take place. 
Therefore, $R_{\bar{\Lambda}}$ looks to be a good observable to detect the vortex rings. 

\vspace*{-8mm}

\begin{acknowledgments} 
\vspace*{-3mm}
Fruitful discussions with E. E. Kolomeitsev, O. V. Teryaev and D. N. Voskresensky are gratefully acknowledged.
This work was carried out using computing resources of the federal collective usage center ``Complex for simulation and data processing for mega-science facilities'' at NRC "Kurchatov Institute" \cite{ckp.nrcki.ru}.
This work was partially supported by MEPhI within the Federal Program ”Priority-2030”.
\end{acknowledgments}


\begin{thebibliography}{999}
%
\bibitem{Ivanov:2018eej}
Y.~B.~Ivanov and A.~A.~Soldatov,
``Vortex rings in fragmentation regions in heavy-ion collisions at $\sqrt{s_{NN}}=$ 39 GeV,''
Phys. Rev. C \textbf{97}, no.4, 044915 (2018)
[arXiv:1803.01525 [nucl-th]].
%
\bibitem{Xia:2018tes}
X.~L.~Xia, H.~Li, Z.~B.~Tang and Q.~Wang,
``Probing vorticity structure in heavy-ion collisions by local $\Lambda$ polarization,''
Phys. Rev. C \textbf{98}, 024905 (2018)
[arXiv:1803.00867 [nucl-th]].
%
\bibitem{Wei:2018zfb}
D.~X.~Wei, W.~T.~Deng and X.~G.~Huang,
``Thermal vorticity and spin polarization in heavy-ion collisions,''
Phys. Rev. C \textbf{99}, no.1, 014905 (2019)
[arXiv:1810.00151 [nucl-th]].
%
\bibitem{Ivanov:2019wzg}
Y.~B.~Ivanov, V.~D.~Toneev and A.~A.~Soldatov,
``Vorticity and Particle Polarization in Relativistic Heavy-Ion Collisions,''
Phys. Atom. Nucl. \textbf{83}, no.2, 179-187 (2020)
[arXiv:1910.01332 [nucl-th]].
%
\bibitem{Zinchenko:2020bbc}
A.~Zinchenko, A.~Sorin, O.~Teryaev and M.~Baznat,
``Vorticity structure and polarization of $\Lambda$ hyperons in heavy-ion collisions,''
J. Phys. Conf. Ser. \textbf{1435}, no.1, 012030 (2020). 
%
\bibitem{Baznat:2013zx} 
  M.~Baznat, K.~Gudima, A.~Sorin and O.~Teryaev,
  ``Helicity separation in Heavy-Ion Collisions,''
  Phys.\ Rev.\ C {\bf 88}, no. 6, 061901 (2013)
  [arXiv:1301.7003 [nucl-th]].
%
\bibitem{Baznat:2015eca} 
  M.~I.~Baznat, K.~K.~Gudima, A.~S.~Sorin and O.~V.~Teryaev,
  ``Femto-vortex sheets and hyperon polarization in heavy-ion collisions,''
  Phys.\ Rev.\ C {\bf 93}, no. 3, 031902 (2016)
  [arXiv:1507.04652 [nucl-th]]. 
%
\bibitem{Tsegelnik:2022eoz}
N.~S.~Tsegelnik, E.~E.~Kolomeitsev and V.~Voronyuk,
``Helicity and vorticity in heavy-ion collisions at NICA energies,''
arXiv:2211.09219 [nucl-th].
%
\bibitem{Lisa:2021zkj}
M.~A.~Lisa, J.~G.~P.~Barbon, D.~D.~Chinellato, W.~M.~Serenone, C.~Shen, J.~Takahashi and G.~Torrieri,
``Vortex rings from high energy central p+A collisions,''
Phys. Rev. C \textbf{104}, no.1, 011901 (2021)
[arXiv:2101.10872 [hep-ph]].
%
\bibitem{Serenone:2021zef}
W.~M.~Serenone, J.~G.~P.~Barbon, D.~D.~Chinellato, M.~A.~Lisa, C.~Shen, J.~Takahashi and G.~Torrieri,
``\ensuremath{\Lambda} polarization from thermalized jet energy,''
Phys. Lett. B \textbf{820}, 136500 (2021)
[arXiv:2102.11919 [hep-ph]].
%
\bibitem{COSY-TOF:2016vhv}
F.~Hauenstein \textit{et al.} [COSY-TOF],
``Measurement of polarization observables of the associated strangeness production in proton proton interactions,''
Eur. Phys. J. A \textbf{52}, no.11, 337 (2016)
[arXiv:1607.06305 [nucl-ex]].
%
\bibitem{HADES:2014ttv}
G.~Agakishiev \textit{et al.} [HADES],
``Lambda hyperon production and polarization in collisions of p(3.5 GeV)+Nb,''
Eur. Phys. J. A \textbf{50}, 81 (2014)
[arXiv:1404.3014 [nucl-ex]].
%
%
\bibitem{3FD}
Y.~B.~Ivanov, V.~N.~Russkikh and V.~D.~Toneev,
``Relativistic heavy-ion collisions within 3-fluid hydrodynamics: Hadronic scenario,''
Phys. Rev. C \textbf{73}, 044904 (2006)
[arXiv:nucl-th/0503088 [nucl-th]].
%
%
\bibitem{Toneev06}
A.~S.~Khvorostukin, V.~V.~Skokov, V.~D.~Toneev and K.~Redlich,
``Lattice QCD constraints on the nuclear equation of state,''
Eur. Phys. J. C \textbf{48}, 531-543 (2006)
[arXiv:nucl-th/0605069 [nucl-th]].
%
\bibitem{Ivanov:2013wha} 
  Yu.~B.~Ivanov,
  ``Alternative Scenarios of Relativistic Heavy-Ion Collisions: I. Baryon Stopping,''
  Phys. Rev. C {\bf 87}, 064904 (2013) [arXiv:1302.5766 [nucl-th]]. 
%
\bibitem{Becattini:2013fla} 
  F.~Becattini, V.~Chandra, L.~Del Zanna and E.~Grossi,
  ``Relativistic distribution function for particles with spin at local thermodynamical equilibrium,''
  Annals Phys.\  {\bf 338}, 32 (2013)
  [arXiv:1303.3431 [nucl-th]].
%
%
\bibitem{Ivanov:2022ble}
Y.~B.~Ivanov and A.~A.~Soldatov,
``Global \ensuremath{\Lambda} polarization in heavy-ion collisions at energies 2.4\textendash{}7.7 GeV: Effect of meson-field interaction,''
Phys. Rev. C \textbf{105}, no.3, 034915 (2022)
[arXiv:2201.04527 [nucl-th]].
%
\bibitem{Russkikh:2006aa} 
  V.~N.~Russkikh and Yu.~B.~Ivanov,
  ``Dynamical freeze-out in 3-fluid hydrodynamics,''
  Phys.\ Rev.\ C {\bf 76}, 054907 (2007)  [nucl-th/0611094];
%
  Yu.~B.~Ivanov and V.~N.~Russkikh,
  ``On freeze-out problem in relativistic hydrodynamics,''
  Phys.\ Atom.\ Nucl.\  {\bf 72}, 1238 (2009)  [arXiv:0810.2262 [nucl-th]].
%
\bibitem{STAR:2017sal}
L.~Adamczyk \textit{et al.} [STAR],
`Bulk Properties of the Medium Produced in Relativistic Heavy-Ion Collisions from the Beam Energy Scan Program,''
Phys. Rev. C \textbf{96}, no.4, 044904 (2017)
[arXiv:1701.07065 [nucl-ex]].
%
\bibitem{Ivanov:2012bh}
Y.~B.~Ivanov,
``Baryon Stopping as a Probe of Deconfinement Onset in Relativistic Heavy-Ion Collisions,''
Phys. Lett. B \textbf{721}, 123-130 (2013)
[arXiv:1211.2579 [hep-ph]];
%
Y.~B.~Ivanov and D.~Blaschke,
``Robustness of the Baryon-Stopping Signal for the Onset of Deconfinement in Relativistic Heavy-Ion Collisions,''
Phys. Rev. C \textbf{92}, no.2, 024916 (2015)
[arXiv:1504.03992 [nucl-th]].
%
\bibitem{Ivanov:2020udj}
Y.~B.~Ivanov,
``Global $\Lambda$ polarization in moderately relativistic nuclear collisions,''
Phys. Rev. C \textbf{103}, no.3, L031903 (2021)
[arXiv:2012.07597 [nucl-th]].
%
\bibitem{Ivanov:2013yqa}
Y.~B.~Ivanov,
``Alternative Scenarios of Relativistic Heavy-Ion Collisions: II. Particle Production,''
Phys. Rev. C \textbf{87}, no.6, 064905 (2013)
[arXiv:1304.1638 [nucl-th]];
%
``Alternative Scenarios of Relativistic Heavy-Ion Collisions: III. Transverse Momentum Spectra,''
Phys. Rev. C \textbf{89}, no.2, 024903 (2014)
[arXiv:1311.0109 [nucl-th]].
%
\bibitem{Ivanov:2014ioa}
Y.~B.~Ivanov and A.~A.~Soldatov,
``Directed flow indicates a cross-over deconfinement transition in relativistic nuclear collisions,''
Phys. Rev. C \textbf{91}, no.2, 024915 (2015)
[arXiv:1412.1669 [nucl-th]].
%
\bibitem{Ivanov:2014zqa}
Y.~B.~Ivanov and A.~A.~Soldatov,
``Elliptic Flow in Heavy-Ion Collisions at Energies $\sqrt{s_{NN}}=$ 2.7-39 GeV,''
Phys. Rev. C \textbf{91}, no.2, 024914 (2015)
[arXiv:1401.2265 [nucl-th]].
%
\bibitem{Panagiotou:1986zq}
A.~D.~Panagiotou,
``$\Lambda^0$ Nonpolarization: Possible Signature of Quark Matter,''
Phys. Rev. C \textbf{33}, 1999-2002 (1986).
%
\bibitem{Ayala:2001jp}
A.~Ayala, E.~Cuautle, G.~Herrera and L.~M.~Montano,
``$\Lambda^0$ polarization as a probe for production of deconfined matter in ultrarelativistic heavy ion collisions,''
Phys. Rev. C \textbf{65}, 024902 (2002)
[arXiv:nucl-th/0110027 [nucl-th]].
%
\bibitem{Bellwied:2002cz}
R.~Bellwied [E896],
``Transverse polarization of Lambda hyperons in relativistic Au - Au collisions at the AGS,''
Acta Phys. Hung. A \textbf{15}, 437-444 (2002). 
%
\bibitem{Okubo:2021dbt}
K.~Okubo [STAR],
``Measurement of global polarization of \ensuremath{\Lambda} hyperons in Au+Au \ensuremath{\sqrt{}}SNN = 7.2 GeV fixed target collisions at RHIC-STAR experiment,''
EPJ Web Conf. \textbf{259}, 06003 (2022)
[arXiv:2108.10012 [nucl-ex]].
%
\bibitem{STAR:2021beb}
M.~S.~Abdallah \textit{et al.} [STAR],
``Global $\Lambda$-hyperon polarization in Au+Au collisions at $\sqrt {s_{NN}}$=3~GeV,''
Phys. Rev. C \textbf{104}, no.6, L061901 (2021)
[arXiv:2108.00044 [nucl-ex]].
%
\bibitem{Adams:2021idn}
J.~R.~Adams [STAR],
``Differential measurements of \ensuremath{\Lambda} polarization in Au+Au collisions and a search for the magnetic field by STAR,''
Nucl. Phys. A \textbf{1005}, 121864 (2021)
%
\bibitem{Nazarova:2021lry}
E.~Nazarova, R.~Akhat, M.~Baznat, O.~Teryaev and A.~Zinchenko,
``Monte Carlo Study of $\Lambda$ Polarization at MPD,''
Phys. Part. Nucl. Lett. \textbf{18} (2021) no.4, 429-438.
%
\bibitem{ATLAS:2014ona}
G.~Aad \textit{et al.} [ATLAS],
``Measurement of the transverse polarization of $\Lambda$ and $\bar{\Lambda}$ hyperons produced in proton-proton collisions at $\sqrt{s}=7$ TeV using the ATLAS detector,''
Phys. Rev. D \textbf{91}, no.3, 032004 (2015)
[arXiv:1412.1692 [hep-ex]].
%
%
\bibitem{ckp.nrcki.ru}
http://ckp.nrcki.ru/
%
\end{thebibliography}
\end{document}